\documentstyle[aps,epsf,floats]{revtex}

\ifx\epsffile\undefined
\def\insertfig#1{}
\else
\def\insertfig#1{{\baselineskip=4pt
\centerline{\epsfxsize=\hsize\epsffile{#1}}}}\fi

\begin{document}
\twocolumn[       
{\tighten

\title{Ghost poles and chiral symmetry in ${\bbox{\pi}}{N}$ scattering\footnotemark}
\author{C.A. da Rocha\footnotemark}
\address{Department of Physics, Box 351560, University of Washington, Seattle, 
Washington 98195-1560}
\author{G. Krein\footnotemark}
\address{Instituto de F\'{\i}sica Te\'orica - Universidade Estadual Paulista\\
Rua Pamplona 145, 01405-900 S\~ao Paulo-SP, Brazil}
\author{L. Wilets\footnotemark}
\address{Department of Physics, Box 351560, University of Washington, Seattle, 
Washington 98195-1560}
\bigskip
\date{May 96}
\maketitle
\widetext
\vskip-2.7in
\rightline{\fbox{
\hbox{DOE/ER/40427-29-N96}
}}

\vskip2.3in
\begin{abstract}
We solve the Schwinger-Dyson equation for the nucleon propagator in the
vacuum using $\pi$, $\sigma$, and $\omega$ mesons. For bare interaction 
vertices (Hartree-Fock approximation) we obtain a pair of complex-conjugated
poles (ghost poles) and for vertices dressed by phenomenological form-factors
these ghosts disappeared. We use these two different approaches to evaluate
the scattering lengths and the phase shifts for the $\pi N$ scattering at 
threshold. Our results show that only when the form factors are present is
possible to obtain a good agreement with the low-energy observables.
\end{abstract}

}

] 
\narrowtext

\footnotetext{${}^*$Contribution presented at
the $14^{\text{th}}$ International Conference on Particles and Nuclei 
(PANIC 96), Williamsburg, VA, 22-28 May 1996. }
\footnotetext{${}^\dagger$Fellow from CNPq, Brazilian Agency.
Electronic address: carocha@phys.washington.edu}
\footnotetext{${}^\dagger$Electronic address: gkrein@axp.ift.unesp.br}
\footnotetext{${}^\ddagger$Electronic address: wilets@nuc2.phys.washington.edu} 


\section{INTRODUCTION}
\label{sec:introduction}
\vspace{0.3cm}
In recent years a renewed interest in $\pi N$ scattering
is being witnessed in the literature~\cite{PJ91,GS93,GLM93}, 
driven mainly by the necessity of having a relativistic description
of the available high energy data, as well as of the data to be generated at
CEBAF. Also, the recognition of chiral symmetry as a fundamental symmetry of 
the strong interactions has motivated a great deal of attention to the role of
this symmetry in hadronic processes. The mathematical complexities presented 
by QCD at low energies require us to use models with effective 
degrees of freedom as, for instance, the relativistic quantum field 
model with baryon and meson degrees of freedom, which we adopted in this work.
In this approach, to go beyond the perturbative scheme, one must evaluate the 
nucleon self-energy by solving the Schwinger-Dyson equation (SDE) and obtain a 
``dressed'' nucleon propagator. The simplest first step calculation is to 
consider ``bare'' vertex interaction and bare meson propagators, which is known 
as Hartree-Fock approximation, and use the low-lying mesons in the SDE
\cite{BPW70,Wil78}.
Using the $\pi$, $\rho$ and $\omega$ mesons,  
the main result of this first approach is the appearance 
of complex poles, or {\em ghost poles}, in the renormalized nucleon propagator.
These ghosts violate basic theorems of quantum field theory and their 
origin is related to the ultraviolet behavior of the model interactions. 
The ghosts disappear when form factors that soften the interaction sufficiently 
in the ultraviolet are used~\cite{Kre+93}. Since these previous approaches
did not consider the role of chiral symmetry, one interesting point is to know
how this symmetry can affect the nucleon self-energies and the 
related observables.

In this work we investigated the $\pi N$ scattering using a linear realization 
of chiral symmetry, which is implemented by the linear $\sigma$-model 
lagrangian  augmented with the $\omega$ meson~\cite{Lee72}. 
The nucleon propagator is obtained by  solving the SDE by the iteration
method. We use this propagator to evaluate the $\pi N$ observables at
threshold in two different situations: with bare interaction vertices and with
vertices dressed by phenomenological formfactors. 

\section{The nucleon spectral function}

\begin{figure}
\vspace{0.5cm}
\epsfxsize=8.0cm
\centerline{\epsffile{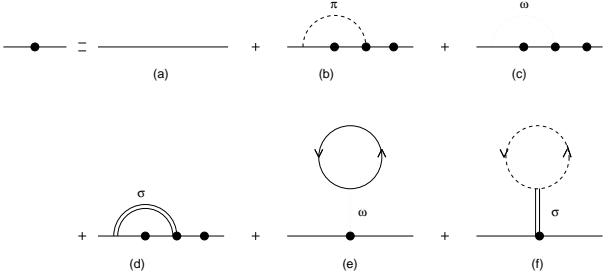}}
\vspace{0.5cm}
\caption{Diagrammatic representation of SDE. The tadpole diagrams do not
contribute because they drop out in the renormalization process.}
\label{Fig.2}
\end{figure}

In this paper we use the spectral representation of the nucleon propagator 
and of its inverse.  We refer the reader to Refs.~\cite{Sch62,Rom69,BLT75} for
more details about this approach. The nucleon propagator is defined as
\begin{equation}
G_{\alpha \beta}(x'-x)=-i<0|T[\psi_{\alpha}(x')\bar \psi_{\beta}(x)]|0>\;,
\label{defnucpro}
\end{equation}
where $|0\!>$ represents the physical vacuum state. 
The K\"allen-Lehmann representation for the Fourier
transform $G(p)$ of $G(x'-x)$ can be written as
\begin{equation}
G(p)=\int_{- \infty}^{+ \infty} d\kappa\; {A(\kappa) \over {{\not\!p} - \kappa
+ i\epsilon}}\;,
\label{kaleh} 
\end{equation}
where $A(\kappa)$ is the spectral function.  It represents the probability that
a state of mass $|\kappa|$ is created by $\psi$ or $\bar \psi$, and as such it
must be non-negative. Negative $\kappa$ corresponds to states with opposite
parity to the nucleon.
The SDE for the nucleon propagator  in momentum 
space is given by the following expressions
\begin{equation}
G(p)=G^{(0)}(p)+G^{(0)}(p)\Sigma(p)G(p),
\label{sdenuc}
\end{equation}
where
\begin{eqnarray}
\Sigma(p)&=&- i g^2_0 \int {d^4q\over (2\pi)^4} \gamma_5 \tau^i
D_{\pi}(q^2)G(p-q)\Gamma_5^i(p-q,p;q) \nonumber \\ [0.2cm]
&+&i g_{0\omega}^2 \int {d^4q\over (2\pi)^4} \gamma_{\mu}
D_{\omega}^{\mu \nu}(q^2)G(p-q)\Gamma_{\nu}(p-q,p;q)\nonumber \\ [0.2cm]
&+&i g^2_0 \int {d^4q\over (2\pi)^4} D_{\sigma}(q^2)G(p-q)\Gamma_{S}(p-q,p;q),
\label{nuceq}
\end{eqnarray}

\noindent
is the nucleon self-energy, shown schematically in Fig.~\ref{Fig.2}. 


In
Eq.~(\ref{nuceq}), $D_\pi$, $D_\omega^{\mu\nu}$, and $D_\sigma$ are the $\pi$, 
$\omega$, and $\sigma$ propagators and $\Gamma_5^i(p-q,p;q)$, 
$\Gamma_\nu (p-q,p;q)$, and $\Gamma_S(p-q,p;q)$  are respectively the
pion-nucleon, omega-nucleon, and sigma-nucleon vertex functions.
The tadpoles in Fig.~\ref{Fig.2} do not contribute to the nucleon propagator,
since they drop out in the renormalization procedure. 
The Hartree-Fock (HF) approximation amounts to using the noninteracting 
meson propagators and bare vertices $\Gamma_5^j(p-q,p;q)=\tau^i\gamma_5$,
$\Gamma_\nu (p-q,p;q)=\gamma_\nu$, and $\Gamma_S(p-q,p;q)=1$ 
in Eq.~(\ref{nuceq}). In order to regulate the ultraviolet behavior of 
the interaction and study the role of a ghost-free self-consistent propagator,
we consider simplified vertex functions that are written as:

\begin{eqnarray}
\Gamma_5^i(p_1, p_2; q)&=&\tau^i \gamma_5 F_5(p_1, p_2; q) 
\label{Gam5} \\ [0.3cm]
\Gamma^{\mu}(p_1, p_2; q)&=&\gamma^{\mu} F_V(p_1, p_2; q) 
\label{GamV}\\ [0.3cm]
\Gamma_S(p_1,p_2;q)&=&F_S(p_1,p_2;q) \label{GamS}\;,
\end{eqnarray}                                               
where $F_5(p_1, p_2; q)$, $F_V(p_1, p_2; q)$, and $F_S(p_1,p_2;q)$
are scalar functions. Next we solve the SDE by iteration~\cite{Kre+93}
and we obtain:

\begin{equation}
T_R(\kappa)=\int_{-\infty}^{+\infty} d\kappa'K(\kappa,\kappa')A_R(\kappa')\;,
\label{tkap}
\end{equation}
where $K(\kappa,\kappa')$ is the scattering kernel given by
\begin{eqnarray}
K(\kappa, \kappa') &=& K_{\pi}(\kappa, \kappa'; m_{\pi}^2) \nonumber \\ [0.2cm]
&+&2 K_{\omega}(\kappa, \kappa'; m_{\omega}^2) +
K_{\sigma}(\kappa, \kappa'; m_{\sigma}^2)\;;
\label{K}
\end{eqnarray}
with $K_{\pi}(\kappa, \kappa'; m^2_\pi)$, $K_{\omega}(\kappa, \kappa'; 
m^2_\omega)$, and $K_{\sigma}(\kappa, \kappa'; m^2_\sigma)$ being
respectively the $\pi$, $\omega$, and $\sigma$ contributions, given by
\begin{eqnarray}
&&K_{\pi}(\kappa, \kappa'; m^2_\pi )= F_5(\kappa, \kappa'; m_\pi) \; 3\left(
{g\over 4\pi}\right)^2\;{1\over 2|\kappa|^3} \nonumber \\ [0.2cm]
&\times&\left[(\kappa-\kappa')^2-m^2_\pi \right]
\theta(\kappa^2-(|\kappa'|+m_\pi)^2) \nonumber \\ [0.2cm]
&\times&\left[\kappa^4-2\kappa^2({\kappa'}^2+m^2_\pi)+
({\kappa'}^2-m^2_\pi)^2\right]^{1/2} \; ,
\label{Kpi}
\end{eqnarray}
\begin{eqnarray}
&&K_{\omega}(\kappa, \kappa'; m_\omega^2)=F_V(\kappa, \kappa'; m_\omega)
\;\left({g_{\omega}\over 4\pi}\right)^2\;{1\over 2|\kappa|^3}\
\nonumber \\ [0.2cm]
&\times&\left[(\kappa-\kappa')^2-2\kappa \kappa'-m_\omega^2\right]
\theta(\kappa^2-(|\kappa'|+m_\omega)^2)\nonumber \\ [0.2cm]
&\times&\left[\kappa^4-2\kappa^2({\kappa'}^2+m_\omega^2)+({\kappa'}^2
-m_\omega^2)^2\right]^{1/2}\;,
\label{Komega}
\end{eqnarray}
\noindent and 
\begin{eqnarray}
&&K_{\sigma}(\kappa, \kappa'; m^2_\sigma )=F_S(\kappa, \kappa'; m_\sigma)
\;\left({g\over 4\pi}\right)^2\;{1\over 2|\kappa|^3}
\nonumber \\ [0.2cm]
&\times&\left[(\kappa+\kappa')^2-m^2_\sigma \right]
\theta(\kappa^2-(|\kappa'|+m_\sigma)^2)\nonumber \\ [0.2cm]
&\times&\left[\kappa^4-2\kappa^2({\kappa'}^2+m^2_\sigma )+({\kappa'}^2
-m^2_\sigma )^2\right]^{1/2}\; .
\label{Ksigma}
\end{eqnarray}
In the above equations, $g$ and $g_\omega$ are the renormalized coupling
constants, defined as $g=Z_2g_0$ and $g_\omega=Z_2 g_{0\omega}$, where 
$Z_2$ is the renormalization constant~\cite{Kre+93}. 
The spectral functions $A_R(\kappa)$ and $T_R(\kappa)$ are related by 

\begin{eqnarray}
A_R(\kappa)&=&\delta(\kappa-M)+
|{\tilde G}_R^{-1}(\kappa(1+i\epsilon))|^{-2}T_R(\kappa)
\label{AandT} \\ [0.2cm]
&\equiv &\delta(\kappa-M)+\bar A_R(\kappa)\;.
\label{Abar}
\end{eqnarray}
\noindent where 
\begin{eqnarray}
&&\tilde G_{R}^{-1}(z) = (z-M) \times\nonumber \\ [0.3cm]
&&\left[ 1-(z-M)\int_{- \infty}^{+\infty} d\kappa\;
{T_{R}(\kappa) \over {(\kappa-M)^2(z-\kappa)}}\right]\;.
\label{rself}
\end{eqnarray}

Initially, we consider bare vertices: $F_5(p_1, p_2, q)=F_V(p_1, p_2, q)=
F_S(p_1, p_2,q)=1$, and study the convergence properties of the SDE for the
nucleon self-energy.  We use the
following values for the coupling constants:
\begin{equation}
{g^2_{\pi}\over 4\pi}=\frac{g^2_\sigma}{4\pi} \equiv \frac{g^2}{4\pi} =
14.6 \hspace{0.4cm}\text{ and }\hspace{0.4cm}
 {g_{\omega}^2 \over 4\pi}=6.36 \;,
\label{couplings}
\end{equation}
where we wrote explicitly that the value of the sigma-nucleon coupling constant
is equal to the pion-nucleon one, as required by the linear realization of
chiral symmetry.

\begin{figure}
\epsfxsize=6.8cm
\epsfysize=8.8cm
\centerline{\epsffile{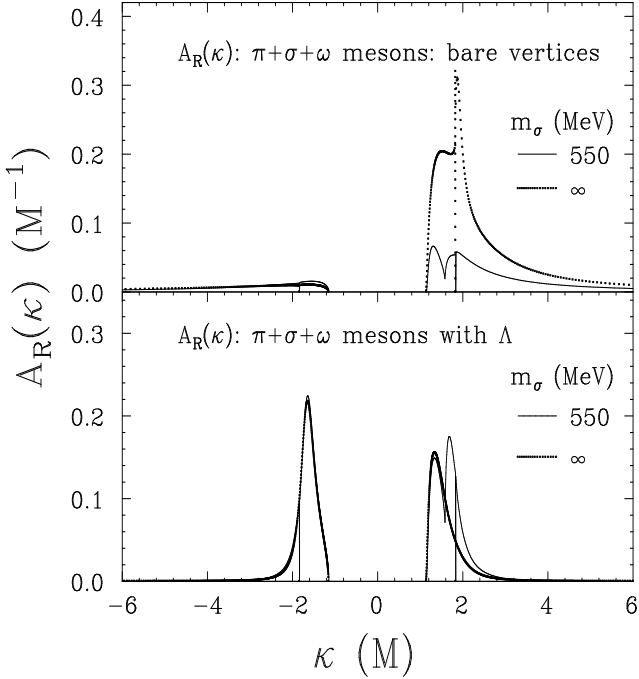}}
\vspace{1.0cm}
\caption{{\it Top}: plot of $A_R(\kappa)$ as a function of $\kappa$
for nucleon self-energy due to $\pi+\sigma+\omega$ mesons with {\em bare}
nucleon vertices. {\it Bottom}: the same study for the same system but with 
vertices dressed by form factors.}
\label{Fig.4}
\end{figure}

We first solved the SDE for bare interaction vertices and we observed
that the introduction of the
chiral partner of the pion {\em does not remove} the ghost poles.
As the mass of the $\sigma$ meson remains a point of debate, we varied 
$m_\sigma$ over a wide range. The solutions of SDE converge quickly in the
studied range, $500\leq m_\sigma \leq 1500$ MeV.  The converged spectral 
functions $A_R(\kappa)$ are shown
in Fig.~\ref{Fig.4}. We used the following values for the meson masses:
\begin{eqnarray}
m_\pi&=&138.03\text{ MeV}\;, \\ [0.2cm]
m_\omega&=&783\text{ MeV}\;, \\ [0.2cm]
m_\sigma&=&550,\;770,\;980\text{ MeV and }m_\sigma\rightarrow
\infty\;.
\label{msigma}
\end{eqnarray}
The reason for using this particular set of $\sigma$ masses is the following: 
$m_\sigma=550$ MeV is the value commonly used in the One Boson 
Exchange Potentials (OBEP)~\cite{MHE}; $m_\sigma=770$ MeV is the 
value used by  Serot and Walecka~\cite{SW86} in the calculations of nuclear
matter properties using the chiral linear sigma model; $m_\sigma=980$ MeV
is the first scalar meson in the mesons table, $f_0$, and the limit
$m_\sigma\rightarrow\infty$ supplies the connection between the linear
realization of chiral symmetry and the minimal chiral model of the non-linear
realization of chiral symmetry in $\pi N$ system \cite{RR94}.


Fig.~\ref{Fig.4}-top presents the nucleon dressed by the $\pi+\sigma+\omega$
mesons, for $m_\sigma=550$ MeV and $m_\sigma\rightarrow\infty$. We note that 
$A_R(\kappa)$ for positive $\kappa$ is much larger than for negative $\kappa$,
it increases as the $\sigma$ mass increases, and becomes equal to the 
$\pi +\omega$ case  in the limit $m_\sigma\rightarrow\infty$, since 
the $\sigma$ meson does not contribute to $A_R(\kappa)$ in this limit.

Next we consider vertex form factors. In principle one can use the
corresponding Sudakov form factors, as in Ref.~\cite{Kre+93}. However, since
the Sudakov form factor is known exactly in the ultraviolet only, one has
to interpolate it in some way down to the infrared or simply parametrize its
infrared behavior. However, since for our purposes here of killing the
ghosts the infrared behavior is not relevant, we prefer to simplify matters
and use parametrized form factors. For general off-shell legs, we use the
factorized form of Pearce and Jennings~\cite{PJ91}; for a vertex with 
four-momenta $p_\alpha$, $p_\beta$, $p_\gamma$, the form factor is
\begin{equation}
F_{\alpha\beta\gamma}=F_\alpha\left(p_\alpha^2\right)
                      F_\beta \left(p_\beta^2 \right)
                      F_\gamma\left(p_\gamma^2\right)\;.
\end{equation}
For mesons we adopt the expressions by Gross, Van Orden and Holinde~\cite{GVH92}
\begin{equation}
F_m\left(q^2\right)=\left[\frac{1+\left(1-\mu_m^2/\Lambda_m^2\right)^2}
{1+\left(1-q^2/\Lambda_m^2\right)^2}\right]^2
\label{ffm}
\end{equation}
where $\Lambda_m$ is the meson cutoff mass. For the nucleon legs we adopt
the expressions by Gross and Surya~\cite{GS93}
\begin{equation}
F_B\left(p^2\right)=\frac{\left(\Lambda_B^2-m_B^2\right)^2}
               {\left(\Lambda_B^2-m_B^2\right)^2+\left(m_B^2-p^2\right)^2}
\label{ffp}
\end{equation}
\noindent where $\Lambda_B$ is the nucleon cutoff mass. Both meson 
and nucleon form factors have the correct on-shell limit, equal to unity.

At this point it is perhaps convenient to call attention that we use form
factors for regulating the ultraviolet with the only aim of studying the role 
of a ghost-free propagator in $\pi N$ scattering. In principle, the form 
factors are calculable within the model by means of vertex corrections. In 
particular, such vertex corrections must satisfy Ward-Takahashi identities 
that follow from chiral symmetry, and of course our form factors $F_m$ 
and $F_B$, Eq.~(\ref{ffm}), do not satisfy such identities. This interesting 
subject is intended to be pursued in a future work.

The cutoff values $\Lambda_B$ and $\Lambda_m$ are constrained to kill the ghost
poles and  give the best fit to the scattering lengths. 
Fig.~\ref{Fig.4}-bottom presents the case $\pi+\sigma+\omega$, with form 
factors at each vertex. We use $\Lambda_B=1330$ MeV. The shape of $A_R(\kappa)$
depends strongly on the $\sigma$ mass for $\kappa>0$ only. One sees that the
second peak is mainly due to the $\sigma$ meson, it decreases as the $\sigma$
mass increases and disappears in the limit $m_\sigma\rightarrow\infty$. 
The interesting effect due to the form factors is that the spectral function 
for $\kappa < 0$ becomes very large as compared to the case of bare vertices.
This will have serious consequences for the observables of $\pi N$ scattering.

\onecolumn
\section{$\pi N$ scattering}
\label{sec:interplay}

\begin{figure}
\vspace{0.5cm}
\epsfxsize=16.0cm
\epsfysize=4.5cm
\centerline{\epsffile{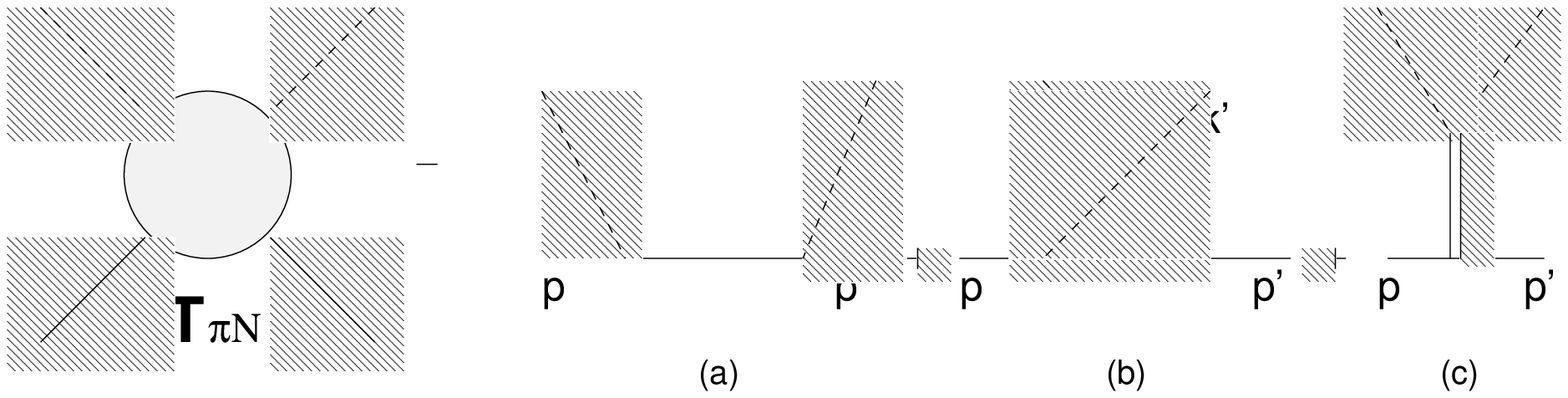}}
\vspace{1.0cm}
\caption{(a),(b) Order $g^2$ contributions involving only pions and
nucleons, where the $\pi NN$ coupling is pseudoscalar.
(c) Scalar meson contribution.}
\label{Fig.6}
\end{figure}

In Born
approximation, when using pions and nucleons only, the lowest order 
contribution is the sum of just two graphs, as shown in Figs.~\ref{Fig.6}(a,b).
It is well known that the first two contributions give bad results for
isoscalar observables when the $\pi NN$ coupling is pseudoscalar.


As we introduce the $\pi\pi\sigma$ and $\sigma NN$ couplings by means of the
chiral symmetry the diagram of Fig.~\ref{Fig.6}(c) appears, which inclusion led
to an almost perfect fit to the value of the isospin-even amplitude $A^{(+)}$
at threshold. This is one of the classical examples of the importance of chiral
symmetry in hadronic interactions. 

The $\pi N$ amplitude $T_{\pi N}$ can be parameterized as~\cite{nutwil}
\begin{equation}
T_{\pi N}=\bar{u}\,({\bf p}')\left\{\left[{A}^+ +
   \frac{1}{2}\,(\not\! k + \not\! k^\prime) B^+\right] 
   \delta_{ab} +\left[A^- +\frac{1}{2}\,
  (\not\! k + \not\! k') 
  B^-\right] i \epsilon_{bac}\, \tau_c\right\}u({\bf p})\;,
\label{tamp}
\end{equation}

\noindent The contribution of the
non-delta part of the spectral function, defined in
Eq.~(\ref{Abar}), to the functions $A^\pm$, $B^\pm$ is given by:
\begin{eqnarray} 
A^\pm(s,t,u)&=&g^2\int_{-\infty}^\infty d\kappa\,\bar A(\kappa)\,
(\kappa-M)\, \left[\frac{1}{s-\kappa^2}\pm\frac{1}{u-\kappa^2}\right]  
\label{A1} \\
B^\pm(s,t,u)&=&g^2\int_{-\infty}^\infty d\kappa\, \bar A(\kappa)\, \left[
-\frac{1}{s-\kappa^2}\pm\frac{1}{u-\kappa^2}\right] 
\label{A3}
\end{eqnarray}
The total contribution for $A^\pm$ and $B^\pm$ is the sum of three parts
coming from: (a) the delta function part of $A(\kappa)$ (nucleon pole term),
(b) the self-energy given by Eqs.~(\ref{A1}-\ref{A3}),
and in the case of bare vertices (c) the ghost poles. The contribution from 
the ghosts poles is given by:
\begin{eqnarray} 
A^\pm_g(s,t,u)&=&g^2\sum_c A_c\, (\kappa_c-M)\,
\left[\frac{1}{s-\kappa_c^2}\pm\frac{1}{u-\kappa_c^2}\right]\label{AC1} \\
B^\pm_g(s,t,u)&=&g^2\sum_c A_c\,\left[-\frac{1}{s-\kappa_c^2} \pm
\frac{1}{u-\kappa_c^2}\right], 
\label{AC2}
\end{eqnarray}
where $\kappa_c$ is the pole position and $A_c$ is the residue and 
the sum is over ($\kappa_c,A_c$) and its complex conjugate
($\kappa_c^*,A_c^*$). 

Tab.~\ref{Tab.05} shows the results for $A^+$ and the scattering lengths 
$a^\pm$ for two cases: $m_\pi=138.03$ MeV and the chiral limit $m_\pi=0$. 
The low-energy theorems impose in the second case that $A^+=1$ 
(in $g^2/M$ units) and $a^\pm=0$. All results are obtained with no form
factors in either the Schwinger-Dyson equation nor in the scattering 
amplitudes.

\begin{table}
\caption{Results for observables using dressed nucleon propagators: (a)
$m_\pi=$138.08 MeV and (b) the chiral limit $m_\pi\rightarrow 0$. No form factors are used.}
\begin{tabular}{l|ccc|rrr} 
             &\multicolumn{3}{c|}{(a) $m_\pi=$138.08 MeV}            
             &\multicolumn{3}{c}{(b) $m_\pi=0$}  \\ \tableline
Contribution & $A^+$  & $a^+$              & $a^-$ & $A^+$ & $a^+$     & $a^-$
\\ 
& ($g^2/M$)& (fm) & (fm)    & ($g^2/M$)       & (fm)   & (fm)   \\ 
\tableline
1) $\pi+\sigma(550)+\omega$: $\bar{A}_R(\kappa)$ 
        &  0.0197  &  0.0773  & 0.1108   &  0.0202   &  0.0619  & 0  \\ 
2) $\pi+\sigma(550)+\omega$: ghosts 
        & -0.5894  & -1.9462  & 1.8702   & -0.7717   & -2.3679  & 0  \\  
3) $\pi+\sigma(550)+\omega$: Tree diagrs.
        &  0.9370  & -0.1830  & 0.1977   &  1.0      &   0      & 0  \\      
Sum (A): $1+2+3$
        &  0.3673  & -2.0520  & 2.1790   &  0.2484   & -2.3606  & 0  \\ 
\tableline
4) $\pi+\sigma(\infty)+\omega$: $\bar{A}_R(\kappa)$
        & -0.1659  & -0.3935  &  0.0891  & -0.2067   & -0.6342  & 0  \\ 
5) $\pi+\sigma(\infty)+\omega$: ghosts
        &  0.9906  &  2.6735  & -0.0821  &  1.0839   &  0.9314  & 0  \\
6) $\pi+\sigma(\infty)+\omega$: Tree diagrs.
        &  1.0     & -0.01453 &  0.1977  &  1.0      &    0     & 0  \\
Sum (B): $10+11+12$
        &  1.8247  &  2.2654  &  0.2074  &  1.8772   &  2.6917  & 0  
\end{tabular}
\label{Tab.05}
\end{table}

The results for the scattering lengths should be compared with the
experimental values given by~\cite{OO75} 
\begin{eqnarray}
a^+&=&-(0.021\pm0.021) \mbox{ fm} \nonumber \\
a^-&=&0.139 \left\{\begin{array}{ll}
                           +0.004 & \mbox{ fm} \\
                           -0.010 & \mbox{ fm}
                                        \end{array}
\right.
\label{Eq.2}
\end{eqnarray}
\noindent and with the predictions of the
low-energy theorems. Inspecting Tab.~\ref{Tab.05}, we see that  
the observable $a^+$ are very far from the experimental result and the 
observables in the chiral limit are poor described.

In order to improve the description of the data, we dressed the interaction
vertices with the formfactors. We fix the cutoff values by fixing the right
values of the low-energy theorems. 
Tab.~\ref{Tab.08} presents the results for the observables for $m_\pi=138.03$
MeV. 


\begin{table}
\caption{Same as in TABLE I, but using form factors.} 
\vspace{0.3cm}
\begin{tabular}{lrrrrrr} 
Contribution    & $A^+$ & $m_\pi\cdot B^+$& $A^-$ & $m_\pi\cdot B^-$& $a^+$ 
                  & $a^-$  \\ 
                & $g^2/M$ & $g^2/M$ & $g^2/M$ & $g^2/M$ & (fm)     & (fm)   \\ 
\tableline
1) $\pi+\sigma(550)+\omega$: $\bar{A}_R(\kappa)$
                & 0.3744  &  0.0249 & 0.1335  & 0.0644  &  1.06820 & 0.5292 \\
2) $\pi+\sigma(550)+\omega$: Tree diagrs.
                & 0.3986  & -0.8335 &   0     & 0.0613  & -1.16336 & 0.1639 \\
Sum (A): $1+2$
                & 0.7730  & -0.8084 & 0.1335  & 0.1267  & -0.09516 & 0.6931 \\ 
\tableline 
3) $\pi+\sigma(\infty)+\omega$: $\bar{A}_R(\kappa)$
                & 0.3815  &  0.0246 & 0.1329  & 0.0601  &  1.08640 & 0.5163 \\
4) $\pi+\sigma(\infty)+\omega$: Tree diagrs.
                & 0.4211  & -0.8335 &    0    & 0.0613  & -1.10326 & 0.1639 \\
Sum (B): $7+8$
                & 0.8026  & -0.8089 & 0.1329  & 0.1214  & -0.01686 & 0.6802 
\end{tabular}
\label{Tab.08}
\end{table}

Observable $A^+$ receives contributions mainly from (a) the
$\sigma$-exchange, Fig.~\ref{Fig.6}(c), which depends directly on $\Lambda_m$,
and (b) from the nucleon self-energy $\bar A_R(\kappa)$, which is weakly
dependent on $\Lambda_B$ and is almost independent from the $\sigma$ mass. 
The results show that $\sigma$ exchange and the self-energy contribution
$\bar A_R(\kappa)$ contribute approximately 50\% each. The results for $a^+$
depend on the sum of $A^+$ and $B^+$. Observable $B^+$ receives contributions
from the nucleon Born part, Fig.~\ref{Fig.6}(a),(b), and from the spectral
function $\bar A_R(\kappa)$. The contribution from the spectral function is
almost constant and very small. The contribution from nucleon pole term 
depends very weakly on $\Lambda_B$. For $\Lambda_B=1.2\rightarrow 1.4$ GeV, 
we obtain $m_\pi\cdot B^+=-0.78\rightarrow -0.86\;g^2/M$. Therefore the way to
get the almost zero result for $a^+$ is decreasing the $A^+$ contribution 
from 1 (without form factor) to 0.8 $g^2/M$, adjusting $\Lambda_m$, since
$m_\pi\cdot B^+\approx -0.8$. 

Observable $a^-$ is not well adjusted due to the huge contribution from
$\bar{A}_R(\kappa)$, being almost 75\% of the final result. This huge 
contribution comes from negative $\kappa$ part of the spectral function. 
We checked this point by doing $\bar{A}_R(\kappa)=0$ for $\kappa<0$ by hand and
get a result 5 times smaller for $a^-$.
As discussed previously, the negative $\kappa$ enhancement is an effect
due to the form factors.

Now we show the phase shifts results.
The total amplitude for $\pi N$ scattering may be decomposed into the 
isospin $\frac{3}{2}$ and $\frac{1}{2}$ channels. The isospin $\frac{3}{2}$
and $\frac{1}{2}$ amplitudes are related to the symmetric and 
antisymmetric amplitudes by

\begin{eqnarray}
O^{(3/2)}&=&O^{(+)}-O^{(-)}\;, \nonumber\\
O^{(1/2)}&=&O^{(+)}+2O^{(-)}\;.
\label{op}
\end{eqnarray}

\noindent
where $O$ can be $A$ or $B$. Therefore, 
the $T$ matrix can be decomposed into good isospin and total angular momentum
channels to reveal the existence of any resonances. 

\begin{figure}
\vspace{0.5cm}
\epsfxsize=7.0cm
\epsfysize=9.0cm
\centerline{\epsffile{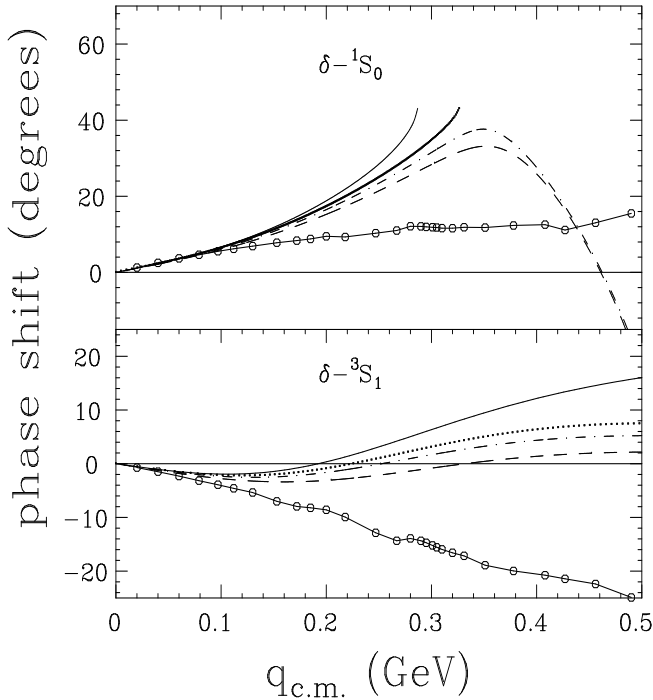}}
\vspace{1.0cm}
\caption{Phase shifts for $\ell=0$ waves. The curves represent the choices
for the $\sigma$-meson mass: 550 MeV (dashed), 770 MeV (dot-dashed), 980 MeV
(dotted), and $m_\sigma\rightarrow\infty$ (solid curve). Data are from
\protect\cite{Hol83}.}
\label{Fig.16}
\end{figure}

The optical theorem identifies the total cross section with the imaginary part
of the scattering amplitude. We unitarize the amplitude following the method
of Olsson and Osypowski~\cite{OO75}.
Fig.~\ref{Fig.16} presents the results for the  phase shifts for $\ell=0$ waves.
The agreement near threshold is very good, and the model fails for 
higher energies. The other phase-shifts show the same trend: they are well 
described near threshold, but as energy increases the agreement with data
becomes poor.

\twocolumn


\section{CONCLUSIONS}
\label{sec:conclusions}

In this contribution we obtained a ``dressed'' nucleon propagator 
by solving the SDE with $\pi$, $\sigma$, and $\omega$ mesons. The 
interaction vertices were extracted from the linear-$\sigma$ model 
augmented with the $\omega$ meson and 
we considered two cases: bare and ``dressed'' vertices, where the 
dressing is made by phenomenological form factors. 
The solution of the SDE for bare vertices contains a pair of ghost poles.
The sum of ghosts plus $\sigma$-meson contributions 
exceeds by a large amount the experimental values of the $a^+$ scattering 
lengths at threshold, 
spoiling the well-known good agreement of the linear $\sigma$-model
for the bare nucleon propagator. The next natural step is to go beyond the 
Hartree-Fock approximation, in order to eliminate the ghosts.
If the ghosts are killed by softening the ultraviolet
by means of form factors it is possible to obtain qualitative agreement with 
experimental data of observables at low energies.  Another lesson from our
study is that the nucleon spectral function $\bar A(\kappa)$ for 
negative $\kappa$ is strongly enhanced by the form factors and this affects 
some of the observables. In particular, the negative $\kappa$ enhancement 
increases the isospin antisymmetric scattering length $a^-$. One possible
solution for this problem is (a) the inclusion of other resonances like $\rho$ 
and $\Delta$ in the tree diagrams and (b) to include the vertices 
$\sigma\pi\pi$ and $\rho\pi\pi$ in the Schwinger-Dyson equation. This is highly
non-trivial because one has to solve a set of coupled integral equation where
the integral is now bidimensional. We intend to tackle this point in a 
near future publication.

\acknowledgements

We thank Prof. M. Robilotta and Prof. T. Cohen for the  helpful conversations 
about $\pi N$ scattering and chiral symmetry. This work was partially suported 
by U.S. Department of Energy. The work of C.A. da Rocha
was supported by  CNPq Brazilian Agency.


\begin{references}
\bibitem{PJ91} 
B.C. Pearce and B.K. Jennings, Nucl. Phys. {\bf A528}, 655 (1991).
\bibitem{GS93}
F. Gross and Y. Surya, Phys. Rev. C {\bf 47}, 703 (1993).
\bibitem{GLM93}
P.F.A. Goudsmit, H.J. Leisi, and E. Matsinos, Phys. Lett. B {\bf 299}, 6 
(1993).
\bibitem{BPW70} 
W.D. Brown, R.D. Puff, and L. Wilets, Phys. Rev. C {\bf 2}, 331 (1970). 
\bibitem{Wil78} 
L. Wilets, in {\it Mesons in Nuclei}, (M. Rho and D. Wilkinson eds., 
North-Holland, Amsterdam, 1979).
\bibitem{Kre+93} 
G. Krein, M. Nielsen, R.D. Puff, and L. Wilets, Phys. Rev. C, {\bf 47} 2485 
(1993); M.E. Bracco, A. Eiras, G. Krein, and L. Wilets, Phys. Rev. C {\bf 49},
1299 (1994).
\bibitem{Lee72} 
B.J. Lee, {\it Chiral Dynamics}, (Gordon and Breach, New York 1972).
\bibitem{Sch62} 
S.S. Schweber, {\it An Introduction to Relativistic Quantum 
Field Theory}, (Harper \& Row, Publ., New York, 1962).
\bibitem{Rom69} 
P. Roman, {\it Introduction to Quantum Field Theory}, (John Wiley \& Sons, 
Inc., New York, 1969).
\bibitem{BLT75}
N.N. Bogolubov, A.A. Logunov, and I.T. Todorov, {\it Axiomatic Quantum Field 
Theory},  (Benjamin, Reading MA, 1975, pp. 269-270, 330ff, Appendix F).
\bibitem{NN}
V.G.J. Stoks, R.A.M. Klomp, M.C.M. Rentmeester, and J.J. de Swart, 
Phys. Rev. C {\bf 48}, 792 (1993).
\bibitem{MHE} 
R. Machleidt, K. Holinde, and Ch. Elster, Phys. Rep. {\bf 149}, 1 (1987).
\bibitem{SW86}
B.D. Serot and J.D. Walecka, {\it Adv. in Nucl. Phys.}, Vol. {\bf 16}, 
(J.W. Negele and E. Vogt eds., Plenum Press, New York, 1986).
\bibitem{RR94} 
C.A. da Rocha and M.R. Robilotta, Phys. Rev. C, {\bf 49}, 1818 (1994); 
{\bf 52}, 531 (1995).
\bibitem{GVH92} 
F. Gross, J.W. Van Orden, and K. Holinde, Phys. Rev. {\bf 45},2094 (1992).
\bibitem{nutwil} 
W.T. Nutt and L. Wilets, Phys. Rev. D {\bf 11}, 110 (1975).
\bibitem{OO75}
M.G. Olsson and E.T. Osypowski, Nucl. Phys. {\bf B101}, 136 (1975).
\bibitem{Hol83} 
G. H\"ohler, Pion-nucleon scattering, {\it in Landolt - B\"ornstein},
Vol. {\bf I/9B-2}, (H. Schopper, ed., Springer Verlag, Heidelberg, 1983).

\end{references}
\end{document}